\documentclass[twocolumn,showpacs,amsmath,amssymb]{revtex4}

\usepackage{graphicx}
\usepackage{dcolumn}
\usepackage{latexsym}
\usepackage{bm}

\newcommand{\srm}[1]{{\textrm{\scriptsize{#1}}}}
\newcommand{\DCI}{\ensuremath{D_\srm{CI}} }

\begin{document}

\thispagestyle{empty}
\title{QCD with two dynamical flavors of chirally improved quarks}
\author{C. B. Lang}%
\email{christian.lang@uni-graz.at}
\author{Pushan Majumdar}%
\email{pushan@uni-muenster.de}
\author{Wolfgang Ortner}%
\email{wolfgang.ortner@uni-graz.at}
\affiliation{{\rm(Bern-Graz-Regensburg (BGR)
collaboration)}\vspace{1mm}\\ Institut f\"ur Physik, FB Theoretische Physik\\
Universit\"at Graz, A-8010 Graz, Austria 
}

\date{February 28, 2006}

\begin{abstract}
Considering Ginsparg-Wilson type fermions dynamically in lattice QCD simulations
is a challenging task. The hope is to be able to approach smaller
pion masses and to eventually reach physical situations. The price to pay is
substantially higher computational costs. Here we discuss first results
of a dynamical implementation of the so-called Chirally Improved Fermions, 
a Dirac operator that obeys the Ginsparg-Wilson condition approximately. The
simulation is for two species of mass-degenerate quarks on $12^3\times 24$
lattices with spatial size up to 1.55 fm. Implementation of
the Hybrid Monte-Carlo algorithm and an analysis
of the results are presented. 
\end{abstract}

\maketitle


\newpage
\section{Introduction}

Quantum Chromodynamics (QCD) is a quantum field theory with a complicated
vacuum structure, involving creation and annihilation of particle-antiparticle
pairs in the vacuum.  Presently the most promising approach for studying this 
theory non-perturbatively seems to be through Monte-Carlo simulations of the 
theory on a space-time lattice. Fermions, however, are described by Grassmann
variables and have no simple representation appropriate for simulations.
Fortunately the fermions occur bilinearly in the action and in the path integral
the Grassmann integration can be performed explicitly. This leads to the fermion
determinant as a weight factor. Each fermion species accounts for one such
factor. 

In lattice QCD Monte-Carlo simulations the path integral is approximated by averages over a
finite set of configurations on a finite  euclidian space-time grid. It is well
understood how to sample gauge field configurations with the full dynamics of
interacting gauge fields alone. It is a challenging task, however, to take into
account the contribution of the fermion determinant in generating the
probability measure. In the quenched approximation these determinants are put to
unity by hand. Here we perform a simulation with full dynamical quarks, i.e.,
taking this determinant into account. 

One of the crucial problems in putting fermions on the lattice is preserving
their chiral symmetry. As an example consider the original Wilson Dirac operator.  It
exhibits spurious zero eigenvalues even at values of the fermionic  mass
parameter that correspond to non-zero pion masses. This feature complicates the
approach to the chiral limit. On the other hand, Dirac operators obeying the
Ginsparg-Wilson (GW) condition  \cite{GiWi82} exactly, have no such spurious
modes.  However, the only known explicit and exact realization of such an
operator, the so-called overlap operator \cite{NaNe93a95,Ne9898a} is quite
expensive to construct and  implement in dynamical simulations
\cite{FoKaSz0405,EgFoKa05,CuKrFr05,CuKrLi05,DeSc05a,DeSc05b}.  Various other
Dirac operators have been suggested, that fulfill the GW condition in an
approximate way, among them the Domain Wall Fermions
\cite{Ka92,FuSh95,AoBlCh04,AnBoBo05}, the Perfect Fermions
\cite{HaNi94,HaHaNi05} or the Chirally Improved (CI) Fermions
\cite{Ga01,GaHiLa00}.  Such Ginsparg-Wilson-type fermions have the chance to
allow a better approach to the chiral limit, i.e., to come closer to the
physical value of the pion mass.

CI fermions have been extensively tested in quenched calculations (see, e.g.,
Ref.~\cite{GaGoHa03a}). In these tests it was found that one can go to smaller
quark masses without running into the problem of exceptional configurations. 
On quenched  configurations pion masses down to 280~MeV could be reached on
lattices of size \(16^3 \times 32\) (lattice spacing 0.148~fm) and about
340~MeV on \(12^3 \times 24\)  lattices.

The challenge is to implement these fermions in dynamical, full QCD
simulations. We discuss here such a simulation for two light, mass-degenerate 
flavors of  dynamical quarks. For updating the gauge fields we use the standard
Hybrid Monte-Carlo (HMC) algorithm. We report on our implementation of \DCI for
the HMC updating and  present results of runs at lattice spacings in the range
0.11--0.13 fm for $12^3\times 24$ lattices of physical spatial extents of  
${\cal O}(1.55\;\textrm{fm})$. Preliminary results on smaller lattices have
been presented elsewhere \cite{LaMaOr05a,LaMaOr05b}.

\section{Algorithmic concerns}
\subsection{The Dirac operator and the action}

The CI Dirac operator \DCI was  constructed by writing a general ansatz for the
Dirac operator
\begin{equation}\label{dcieq}
 D_{ij} = \sum_{k=1}^{16} \alpha^k_{ij}(U)\,\Gamma_k \;,
\end{equation}
where $\Gamma_k$ $(k=1\ldots 16)$ are the 16 elements of the Clifford algebra
and  $\alpha^k_{ij}(U)$ are sums of path ordered products of links $U$
connecting lattice site $i$ with site $j$. Inserting $D$ into the 
GW relation and solving the resulting algebraic equations yields
the CI Dirac operator $\DCI$.  In principle this can be an exact solution, but
that would require an infinite number of terms.  In practice the number of
terms is finite and the operator is a truncated series solution to the
Ginsparg-Wilson relation, respecting the lattice symmetries, parity, invariance
under charge conjugation as well as $\gamma_5$-hermiticity, but connecting
sites only over a certain distance. For our implementation we use terms up to
path lengths of four \cite{Ga01,GaHiLa00}.

Smearing is an essential quality improving ingredient in our Dirac operator. In
the quenched  studies for \DCI \cite{GaGoHa03a} it was found that smearing  the
gauge links was important (using HYP smearing \cite{HaKn01}) as it resulted in
better chiral properties for the  operator, i.e., the spectrum showed less
deviation from the Ginsparg-Wilson circle compared to the unsmeared case.
Therefore we decided to use one level of smearing  in our studies, too. Usual
HYP smearing, however, is not well suited for use in HMC. The recent
introduction of the differentiable ``stout''-smearing \cite{MoPe04} opened the
possibility to implement HMC for smeared fermionic actions. We thus used
one level of isotropic stout smearing of the gauge configuration as part of the
definition of the Dirac operator. Our smearing parameter ($\approx$  0.165
in the notation of Ref.\ \cite{MoPe04}) 
was chosen to maximize the plaquette value. A comparison between  HYP and
stout smearing on loops of different sizes on the same background has been
presented in \cite{LaMaOr05b}.

In quenched calculations \cite{Ga01,GaHiLa00} it was found that the tadpole
improved L\"uscher-Weisz \cite{LuWe85} gauge action had certain
advantages over the Wilson gauge action in the sense that the configurations
produced with  this action were smoother than the ones produced by the Wilson
gauge action. We therefore used this gauge  action in our dynamical
studies. It is given by
\begin{eqnarray}\label{LWaction}
S_g &=&
-\beta_1\sum_\srm{pl}\frac{1}{3}\,\textrm{Re \,\textrm{tr}}\;U_\srm{pl}
-\beta_2\sum_\srm{re}\frac{1}{3}\,\textrm{Re \,\textrm{tr}}\;U_\srm{re}
\nonumber \\
&&
-\beta_3\sum_\srm{tb}\frac{1}{3}\,\textrm{Re \,\textrm{tr}}\;U_\srm{tb}\;,
\end{eqnarray}
where $U_\srm{pl}$ is the usual plaquette term,  $U_\srm{re}$ are Wilson
loops of rectangular $2 \times 1$ shape and  $U_\srm{tb}$ denote loops of
length 6 along edges of 3-cubes (``twisted bent'' or ``twisted chair''). The
coefficient $\beta_1$ is the independent gauge coupling and the other two
coefficients $\beta_2$ and $\beta_3$ are determined from tadpole-improved
perturbation theory. They have to be calculated self-consistently
\cite{AlDiLe95} from
\begin{equation}\label{LWparams}
u_0=\left (\frac{1}{3}\,\textrm{Re \,\ensuremath{\textrm{tr}}}\langle U_\srm{pl}\rangle\right)^{\frac{1}{4}}
\;,\quad
\alpha=-\frac{1}{3.06839}\,\log\left(u_0^4\right)\;,
\end{equation}
as
\begin{equation}
\beta_2=\frac{\beta_1}{20\, u_0^2}\,(1+0.4805\,\alpha)\;,\quad
\beta_3=\frac{\beta_1}{u_0^2}\,0.03325\,\alpha \;.
\end{equation}
This determination should be done for each set of parameters $(\beta_1,m)$.

Putting together all of the above, the partition function that we simulate assumes the form 
\begin{equation}
{\mathcal Z}=\int\,{\mathcal D}U\,{\mathcal D}\phi\,{\mathcal D}\phi^{\dag}\;
e^{-S_g-\phi^{\dag}(M^{\dag}M)^{-1}\phi}\;,
\end{equation}
where $M$ denotes the massive Dirac operator given by
\begin{equation}
M(\overline U,\,m) \equiv\DCI(m)\equiv\DCI(\overline U)+{\bf 1} \,m\;,
\end{equation}
$\overline U$ denotes the smeared link variable, and $\phi$ denotes the 
bosonic pseudofermion field.

\subsection{HMC for CI fermions}

The algorithm we used for our simulations is standard HMC \cite{DuKePe87} as
this  seems to be the most efficient one for simulating fermions at the moment.
Although  our implementation follows  Ref.\ \cite{GoLiTo87} very closely, we
believe that the additional complications due to the  extended structure of our
Dirac operator warrants  some discussion. 

HMC consists of a molecular dynamics (MD) evolution in $2\,n$ dimensional phase
space (where $n$ is the original dimension of the theory) with simulation time
as the evolution parameter. This involves introducing conjugate momenta and
constructing a Hamiltonian for the problem. In our case, to construct a
Hamiltonian, we introduce traceless hermitian matrices $p_{i,\mu}\in su(3)$
acting as momenta conjugate to the \(U_{i,\mu}\) (\(i\) being the site index
and \(\mu\) the direction of the link) and write
\begin{equation}
  \mathcal{H} = \frac{1}{2} \sum_{i,\mu} \mathrm{tr}\left(p_{i,\mu}^2\right) + S_g 
+ \phi^\dagger\left(M^\dagger\, M\right)^{-1}\phi\;.
\end{equation}

There are two evolution equations for $U_{i,\mu}$ and $p_{i,\mu}$.  While
${\dot U}_{i,\mu}=i\,p_{i,\mu}U_{i,\mu}$ , the equation for ${\dot p}_{i,\mu}$
is  obtained by setting $\dot{\mathcal{H}}=0$, 
\begin{equation}
0=\dot{\mathcal{H}}=\sum_{j,\mu} \,\mathrm{tr} \ p_{j,\mu} \,\dot{p}_{j,\mu}  
+ \dot{S_g} +
\phi^{\dag}\,\frac{d}{dt}\left(M^{\dag}M\right )^{-1}\,\phi\;.
\end{equation}
Up to this point there is no difference to HMC with Wilson or staggered
fermions.  The first complication comes when we take the time derivative of our
operator. Unlike Wilson or Staggered Dirac operators, which have only one link
connecting the neighboring sites, we have longer paths. 

As an example let us look at a path of length three.
Let 
\begin{equation}
\mathcal{U}=U_{j_1,\mu_1} U^\dagger_{j_2,\mu_2} U_{j_3,\mu_3}\;.
\end{equation}
Taking the time derivative, we find 
\begin{eqnarray}
\frac{d\,\mathcal{U}}{dt} &=& i\, p_{j_1,\mu_1}\, U_{j_1,\mu_1} U^\dagger_{j_2,\mu_2} U_{j_3,\mu_3} 
\nonumber \\
&&+ U_{j_1,\mu_1}\, U^\dagger_{j_2,\mu_2} (- i\, p_{j_2,\mu_2}) U_{j_3,\mu_3} \nonumber \\
&&+U_{j_1,\mu_1} U^\dagger_{j_2,\mu_2} (i \,p_{j_3,\mu_3} U_{j_3,\mu_3})\;,
\end{eqnarray}
where we have replaced ${\dot U}_{i,\mu}$ by $i\,p_{i,\mu}U_{i,\mu}$. 
The time derivative of the Dirac operator can thus be 
expressed as 
\begin{equation}
\frac{dM}{dt} = \sum_{j,\mu,k} c_{j,\mu,k} 
   W^{(1)}_{j,\mu,k}(\pm i p_{j,\mu}) W^{(2)}_{j,\mu,k}\;,
\end{equation}
where $W^{(1)}$ and $W^{(2)}$ contain products of links. To obtain our equation of
motion for ${\dot p}_{j,\mu}$, we have  to carefully cyclically permute the
terms in $\phi^{\dag}\frac{d}{dt}\left(M^{\dag}M\right)^{-1}\phi$ until all
the  $p_{j,\mu}$ occur at the same position, at the front of the expression.
This is of  course required due to the non-Abelian nature of the variables and
it is possible to cyclically permute them because we have an overall trace over
both color and Dirac indices. In fact the  trace on the Dirac indices can be
carried out completely as the momenta and gauge part of the  Hamiltonian do not
have Dirac indices. The color trace on the other hand should not be carried 
out as we want an equation for the color matrices.

Finally our equation can be symbolically written in the form 
\begin{equation}\label{pdoteq}
p_{j,\mu}\,{\dot p}_{j,\mu}= -i\,p_{j,\mu}\sum (staples)_{j,\mu}
-i\,p_{j,\mu}\sum (Dirac)_{j,\mu}\;.
\end{equation}
From this we get our equation for ${\dot p}_{j,\mu}$.  To ensure the 
group property for the link variables we actually take the traceless 
part of the r.h.s. of (\ref{pdoteq})  for ${\dot p}_{j,\mu}$, as usual. 

Since the $\DCI$ contains several hundred path terms, it was not practical to
do the permutations by hand. Thus the major part of the development stage was
spent in writing  automated routines which perform this task given the table of
coefficients and paths which define  the $\DCI$.

As we have mentioned before, smearing is an important part of the $\DCI$.
The operator is built  not from bare links but from smeared links. The
momenta, however, are conjugate to the bare links. Therefore we have to 
express the smeared links as functions of bare links and compute
corresponding  derivatives.  This is easily done for the stout links and we
will not discuss it further here except to  remark that we used a Taylor
series with forty terms to exponentiate the sum of the staples. While this
is not the most efficient way of doing the exponentiation, it had the
advantage that the differentiation was straightforward and the extra overhead in
time was negligible compared to the time required for inverting the Dirac
operator which is the most time consuming step.   

To sum up some of our operational details, we used one set of pseudofermions
and the leap-frog integration scheme. In order to speed up the conjugate
gradient (CG) inverter, we used a chronological inverter by minimal residual
extrapolation \cite{BrIvLe97}, i.e.,  an optimal linear  combination of the
twelve previous solutions as the starting solution. This of course was used 
only in the MD evolution and reduced the number of iterations required 
to invert \DCI by a factor between  two and three.  For more details on this 
see \cite{LaMaOr05b}. 

\section{Simulation}

We tested our code first on $8^3 \times 16$ lattices \cite{LaMaOr05a,LaMaOr05b}. 
This lattice size does not allow for small bare quark masses, thus we chose $a\,m
= 0.05$ and $0.08$. We then progressed to the larger lattice size $12^3 \times
24$ and smaller quark masses. Table\ \ref{tab:runparameters} summarizes the runs
discussed here.

\begin{table}[bt]
\caption{Parameters for the simulations; the first column denotes the run, 
for later reference.
The gauge coupling is $\beta_1$, the bare quark mass parameter $a m$,
$\Delta t$ the MD time step, $steps$ the number of steps for one trajectory.
In the last three columns we give the acceptance rate in the accept/reject step
(in equilibrium), the total HMC time of the run and the
lattice spacing determined via the Sommer parameter.}
\label{tab:runparameters}
\begin{ruledtabular}
\begin{tabular}{crllllllllll}
\#&    \(L^3 \times T\)  &$\beta_1$  & \(a\,m\)  & $\Delta t$&steps& acc.&HMC  &\(a_S\)[fm]\vspace{-1mm}\\
  &                      &           &           &       &     & rate &time     &          \\
\hline
a&    \(12^3 \times 24\)& 5.2        &  0.02     &0.008  & 120 &0.82(2)  &463      &  0.115(6) \\
b&    \(12^3 \times 24\)& 5.2        &  0.03     &0.01   & 100 &0.94(2)  &363      &  0.125(6) \\
c&    \(12^3 \times 24\)& 5.3        &  0.04     &0.01   & 100 &0.93(1)  &438      &  0.120(4) \\
d&    \(12^3 \times 24\)& 5.3        &  0.05     &0.01   & 100 &0.92(2)  &302      &  0.129(1) \\
e&    \(8^3 \times 16\) & 5.3        &  0.05     &0.015  & 50  &0.93(1)  &1245     &  0.135(3)\\
f&    \(8^3 \times 16\) & 5.4        &  0.05     &0.015  & 50  &0.93(2)  &649      &  0.114(3)\\
g&    \(8^3 \times 16\) & 5.4        &  0.08     &0.015  & 50  &0.94(1)  &776      &  0.138(3)\\
\end{tabular}
\end{ruledtabular}
\end{table}

The parameters of $\DCI$ are fixed by the construction principle discussed in
\cite{GaHiLa00}, i.e., by optimizing the set of algebraic equations that
approximate the Ginsparg Wilson equation. In the formalism there is a
renormalization parameter $z_s$ that has to be adjusted such that $\DCI(m)$ has
the low-lying eigenvalues peaked at $Re(\lambda)=m$. For that one studies a
sample set of gauge configurations  (determined with the full action)
and uses that knowledge to recursively  fix the parameter $z_s$. This recursion
cannot be iterated indefinitely due to limited resources. We therefore decided
to do this adjustment only at one set of parameters $(\beta_1,m) = (5.3,0.05)$
and than hold all couplings fixed (except for $\beta_1$ and $m$). 

The gauge part of the action also requires a recursive approach as the
parameters of the tadpole improved L\"uscher-Weisz gauge action have to be
tuned with help of the plaquette observable \cite{AlDiLe95}. This  can be done
by equating the $u_0^4$ to the moving average of the plaquette defined  over a
suitable time interval. The adjustment should be stopped once equilibrium is 
reached so that the parameters of the action are fixed. Since we started mostly 
from configurations close to equilibrium, we did not go though this whole procedure,
but adjusted the value of $u_0^4$ to be close to its equilibrium value and held 
it fixed. The difference between the assumed plaquette and measured plaquette 
in our runs ranged between 0.2 and 2\%. 

It turned out, that the lattice spacing depends significantly on both, the
gauge coupling and the bare quark mass. The bare parameters of the action have
no real physical significance, however, and we will present most of our results
in terms of derived, physical quantities.

The lattice spacing may be determined in various ways, e.g., from the values of
the measured meson masses or from the static potential. At the distance of the
Sommer parameter \cite{So94}, $r_0=0.5\;\textrm{fm}$ dynamical quark effects are expected
to be small and for the masses discussed here most likely very small. We
therefore use the value of the lattice spacing $a_S$ as derived from the static
potential as our principal scale (cf. Table\ \ref{tab:runparameters}).  For the
determination of the potential we used HYP smeared gauge configurations, since
they showed less fluctuation. Details of this determination are discussed
elsewhere \cite{LaMaOr05a}.

Based on these values for the lattice spacing, our spatial lattice sizes vary
from 1.4 fm up to 1.55 fm (for the $12^3\times 24$ lattices) and we may expect
finite size effects for the derived meson masses. Indeed, meson masses for the
run (e) (small lattice) are about 10\% larger than those on the larger lattice
of run (d). Also in quenched calculations  at comparable parameters the volume
dependence of the meson masses was of a similar size when changing from 
$8^3\times 16$ to $12^3\times 24$ \cite{GaGoHa03a}. Further increase of the
lattice size then  showed significantly smaller changes in the mass values. We
also measured the spatial Wilson lines but observed no signals for
deconfinement.

The CG tolerance values were fixed to $10^{-7}$ in the MD steps and
$10^{-10}$ in the accept/reject step. This followed from earlier experience
on smaller lattices, where we used values down to $10^{-12}$ for both
(see also the discussion in \cite{LaMaOr05a}). 
The number of MD steps and the step sizes are given in Table\
\ref{tab:runparameters}.

\begin{figure}[t]
\includegraphics*[width=8.3cm,clip]{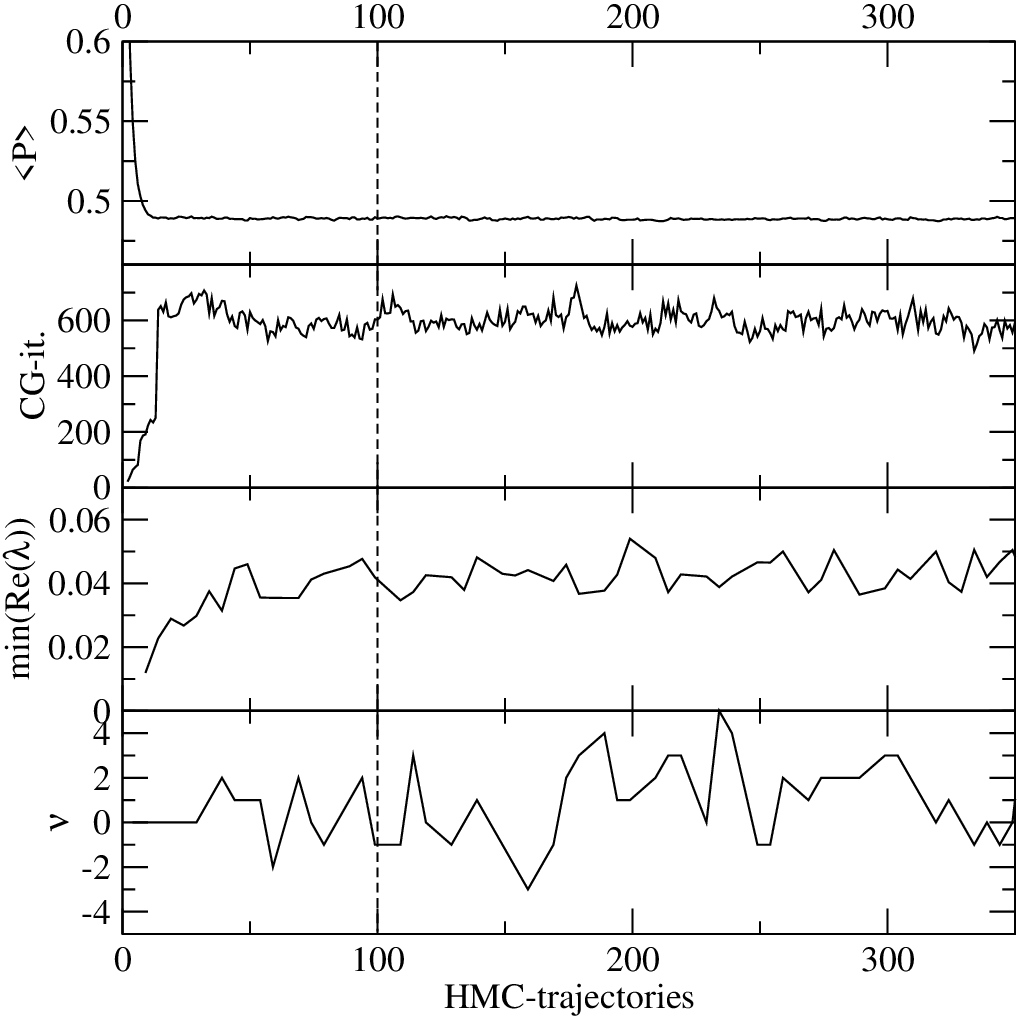}
\caption{
Equilibration signals for $\beta=5.2$, $a\,m=0.03$; from top to bottom:
the plaquette mean, the number of CG iterations for the accept/reject step,
the lower bound $\min(Re(\lambda))$ for the real part of the eigenvalues of the 
Dirac operator $\DCI(m)$, the topological charge $\nu$.
The lower two rows' values have been determined only for every 5th configuration.}
\label{fig:equilibration}
\end{figure}

As an example of the equilibration process we show in Fig.\
\ref{fig:equilibration} some observables computed in the run-sequence for
$\beta_1=5.2$ and  $a\,m=0.03$. This run was started from a cold gauge
configuration and one finds that  the plaquette quickly approaches its
equilibrium value. Another useful, more technical, observable is the number of
CG iterations necessary in the accept/reject step. There equilibrium values are
obtained somewhat later. 

In Fig.\ \ref{fig:equilibration} we also plot the development of the lower
bound to the real parts of the eigenvalues of $\DCI(0.03)$. The value for every
5th configuration is shown. This quantity is a good indicator of the
equilibration as it is most likely the  largest time scale in the problem. At
the same time it indicates the quality of the Dirac operator. Configurations,
where the smallest eigenvalue becomes zero would lead to spurious zero modes.
The amount of fluctuation of the observed lower bound also indicates the
uncertainty width of the choice of the mass.

In the last row in  Fig.\ \ref{fig:equilibration} we show the topological
charge $\nu$, again determined for every 5th configuration. The definition and
determination is discussed in Sect. \ref{sec:eigenvalues}. Starting at a cold
configuration initially locks the topological charge to the trivial sector.
However,  following the initial equilibration period, we find satisfactory
fluctuation.

Altogether, based on such observations, we discarded the first 100 HMC 
trajectories before starting the analyzing measurements. The data are not
sufficient to find reliable estimates for the autocorrelation length. Judging
from the inspection of the fluctuation we estimate autocorrelation times of
10-20. Runs on the smaller lattice size had larger autocorrelation 
times \cite{LaMaOr05a}.
We analyzed every 5th configuration, determining hadron correlators and eigenvalues 
as discussed in Sect.\ \ref{sec:whatdowelearn}.

All our runs were done with MPI parallelization, typically on 8 or 16 nodes on 
the Hitachi SR8000 (at LRZ Munich) or on 8 or 16 nodes of two Opteron 248
processors at 2.2 GhZ per processor. To get an estimate on timing, we note that  for
the $12^3\times 24$ lattice at $a\,m=0.05$ one trajectory takes $\sim 4$ hours
on 8 nodes of the  Hitachi; one trajectory at $a\,m=0.02$ takes
$\sim 7$ hours on 16 of the double Opteron nodes.

\section{What do we learn}\label{sec:whatdowelearn}

The numbers that are presented here are affected by various limitations:
\begin{itemize}
\item We consider only two flavors of mass-degenerate quarks.
\item The typical physical lattice in spatial direction is only up to 1.55 fm. 
We thus may expect to see finite volume effects.
\item The statistics is limited; most of the runs discussed
amount to typically 200 to 350 units of HMC time corresponding to
40 to 70 independent configurations.
\item The smallest quark mass parameters correspond to a pion mass around 500 MeV,
the smallest pion over rho mass ratio is 0.55.
\item Only a small range of lattice spacing values between 0.11 fm and 0.13 fm is covered.
\end{itemize}
All these constraints are typical for first simulations with dynamical quarks,
in particular in view of the computationally demanding properties of the
Dirac operator. We therefore consider this study as the first but necessary step
towards studies on larger and finer lattices with better statistics.

\begin{figure}[t]
\includegraphics*[height=6cm,clip]{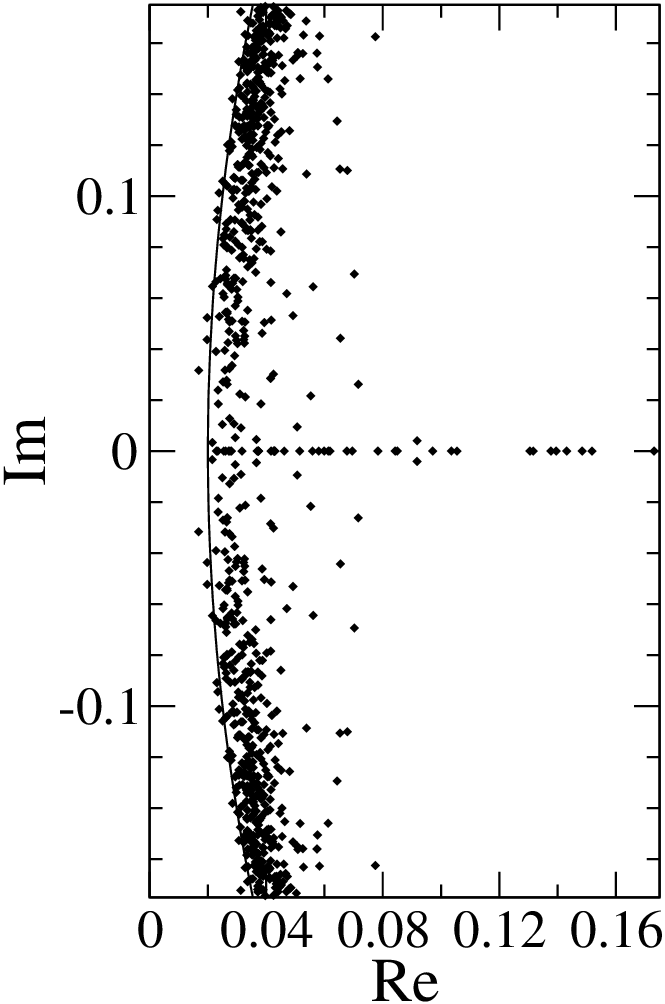}
\caption{
Plot of the 40 smallest eigenvalues of $\DCI(0.02)$ for a total of 21
configuration in equilibrium (run (a), lattice
size $12^3\times 24$,  $\beta_1=5.2$, $a\,m=0.02$).}
\label{fig:eigenvalues}
\end{figure}

\subsection{Eigenvalues and topology}\label{sec:eigenvalues}

Dirac operators obeying the GW condition in its simplest form
have a characteristic eigenvalue spectrum: all
eigenvalues are constrained to lie on a unit circle. The operator $\DCI$ 
is only an approximate
GW operator and its eigenvalues are close to but not exactly on this
so-called GW circle (radius 1, center at 1 in the complex plane).
As a first test we studied the low-lying eigenvalues of $\DCI$ for
some gauge configurations obtained with dynamical quarks.

Fig.\ \ref{fig:eigenvalues} exhibits the accumulated eigenvalues for 21
configurations, where we determined the 40 smallest  eigenvalues for each
configuration. The curve indicates the eigenvalue circle for an exact GW
operator for mass $a\,m=0.02$.  The bulk of the eigenvalues closely follows the
circular shape. A comparison of the spectra on smaller lattices with
the quenched case has been presented in Ref.\ \cite{LaMaOr05a}.

The complex eigenvalues come in complex conjugate pairs  $\lambda_i$ and
$\overline\lambda_i$ and always have vanishing chirality  $\langle \psi_i|
\gamma_5 |\psi_i \rangle$. The real eigenvalues, which would be exact zero modes
for exact GW operators, correspond to topological charges via the
Atiyah-Singer index theorem \cite{AtSi71,HaLaNi98}. For the overlap Dirac
operator only zero modes either all positive or all negative chirality have been
observed. In our case we sometimes (in about 3\% of the configurations) find 
zero modes with opposite chiralities as well.

\begin{figure}[t]
\includegraphics*[width=6.5cm,clip]{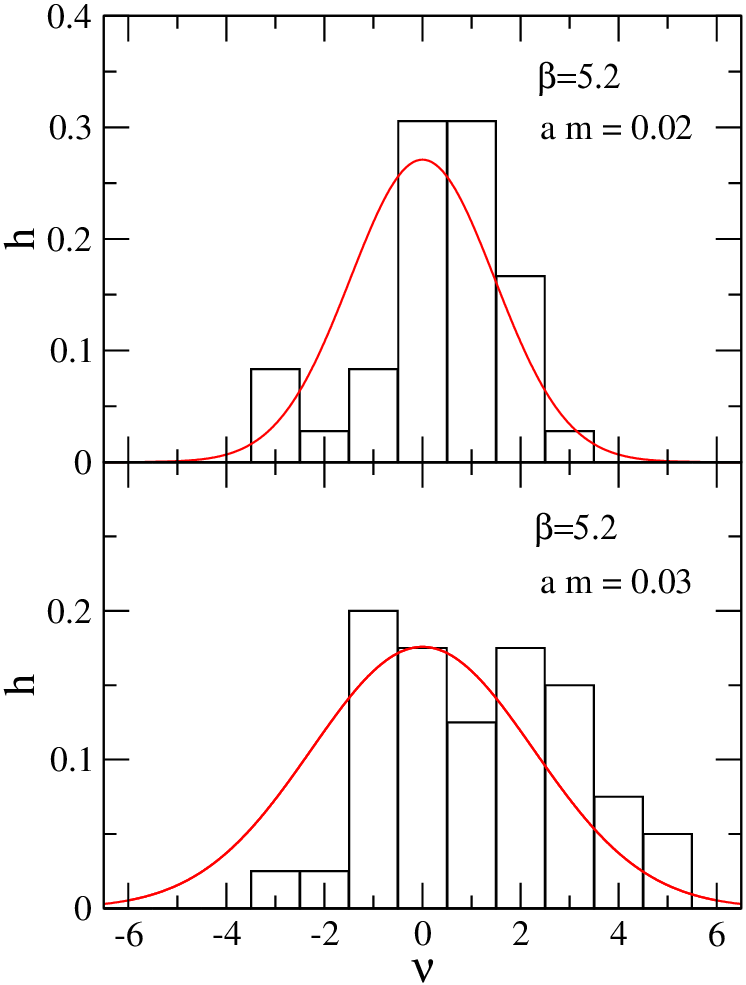}
\caption{
The distribution of the topological charge for the two runs at
$\beta=5.2$ and $\,m=0.02$ and 0.03. We also show a fit to a Gaussian 
distribution function with the same $\langle \nu^2\rangle$.}
\label{fig:topo_hist}
\end{figure}

We did compute the chiralities of the real (zero-)modes for the runs at
$\beta=5.2$. The distribution is not symmetric but this is non-significant due to
the relatively small sample. A symmetrized histogram is in good agreement with
the Gaussian shape. In Fig.\ \ref{fig:topo_hist} we show the resulting histograms together
with a Gaussian distribution with zero mean and the same second moment 
$\langle\nu^2\rangle$. As is well-known from experience with other Dirac operators, when
evaluated over longer HMC-periods than the ones available to us, the
tunneling frequency may show much longer correlation time
\cite{AlBaDe98,AoBlCh04}.

Tunneling between different topological sectors appears to be a problem for  HMC
implementations of the overlap action;
various intricate methods have been
suggested to deal with it
\cite{FoKaSz0405,EgFoKa05,CuKrFr05,CuKrLi05,DeSc05a,DeSc05b}. As can been
seen in the lowest row of Fig.\ \ref{fig:equilibration} we do not seem to 
have such a problem and observe frequent
tunneling.

As mentioned, we determined the eigenvalues and thus the topological charge only
for every 5th configuration. Comparing the number of changes of the topological
sector we get (for the runs (a,b,d) and (e)) between 12 and 19 such
changes along a HMC-time distance of 100, without obvious correlation to the 
run parameters. E.g., runs (d) and (e) have the same values of $\beta_1$ and
$m$, but different  lattice sizes. The width of the distribution of the smaller
lattice is smaller, as expected, but the number of tunneling events comparable.
These values are a lower bound for the actual number of tunneling events.
In view of these we do not find the drastic dependence of the tunneling
rate on the quark mass observed in simulations with the overlap action
\cite{DeSc05b}.

The topological susceptibility
\begin{equation}
\chi_\srm{top}=\langle \nu^2\rangle /V
\end{equation}
plays a central role in the 
Witten-Veneziano formula \cite{Wi79,Ve79,Ve80}.
Due to the axial anomaly the {\em quenched} susceptibility 
(in the large $N_c$-limit) is related
to the $\eta'$-mass via
\begin{equation}
\chi_\srm{top}^{quenched}=\frac{f_\pi^2\,m_{\eta'}^2}{2 N_f}\;.
\end{equation}
(Here $f_\pi\approx 92\;\textrm{MeV}$ denotes the pion decay  constant and $N_f$
the number of flavors, both defined in full QCD.) In quenched simulations values of
$\chi_\srm{top}\approx (190\ \textrm{MeV})^4$ have been found
\cite{GaHoSc02,DePi04,DeGiPi05}.  For the dynamical case the dependence of 
$\chi_\srm{top}$ on $N_f$ and $m$ has been discussed recently, along with a
consistent definition of the quantity \cite{GiRoTe02,Se02,GiRoTe04,Lu04}. Due to
the anomalous Ward identity for the U(1) axial current the topological
susceptibility for full QCD should vanish (in the chiral limit). As in formal
continuum theory \cite{Cr77,LeSm92} the topological susceptibility (now for
dynamical fermions) and the chiral condensate are related via
\begin{equation}
\chi_\srm{top}^{dyn}=-\frac{m\,\Sigma}{N_f} + {\cal O}(m^2)\;.
\end{equation}
where $\Sigma$ denotes the condensate contribution per flavor degree of freedom.

The precision of our results for $\chi_\srm{top}$ is too poor to
verify the linear behavior in the quark mass. For the runs (a) and (b) of Table \ref{tab:runparameters}
we obtain in physical units the values
\begin{eqnarray}
\chi_\srm{top}^{dyn}(a)&=& (146(8) \; \textrm{MeV})^4\;,\nonumber\\
\chi_\srm{top}^{dyn}(b)&=& (166(8) \; \textrm{MeV})^4\;.
\end{eqnarray}
In these cases we do find, as expected, that
our values are definitely smaller than the quenched ones.

\begin{figure}[t]
\includegraphics*[width=8.3cm,clip]{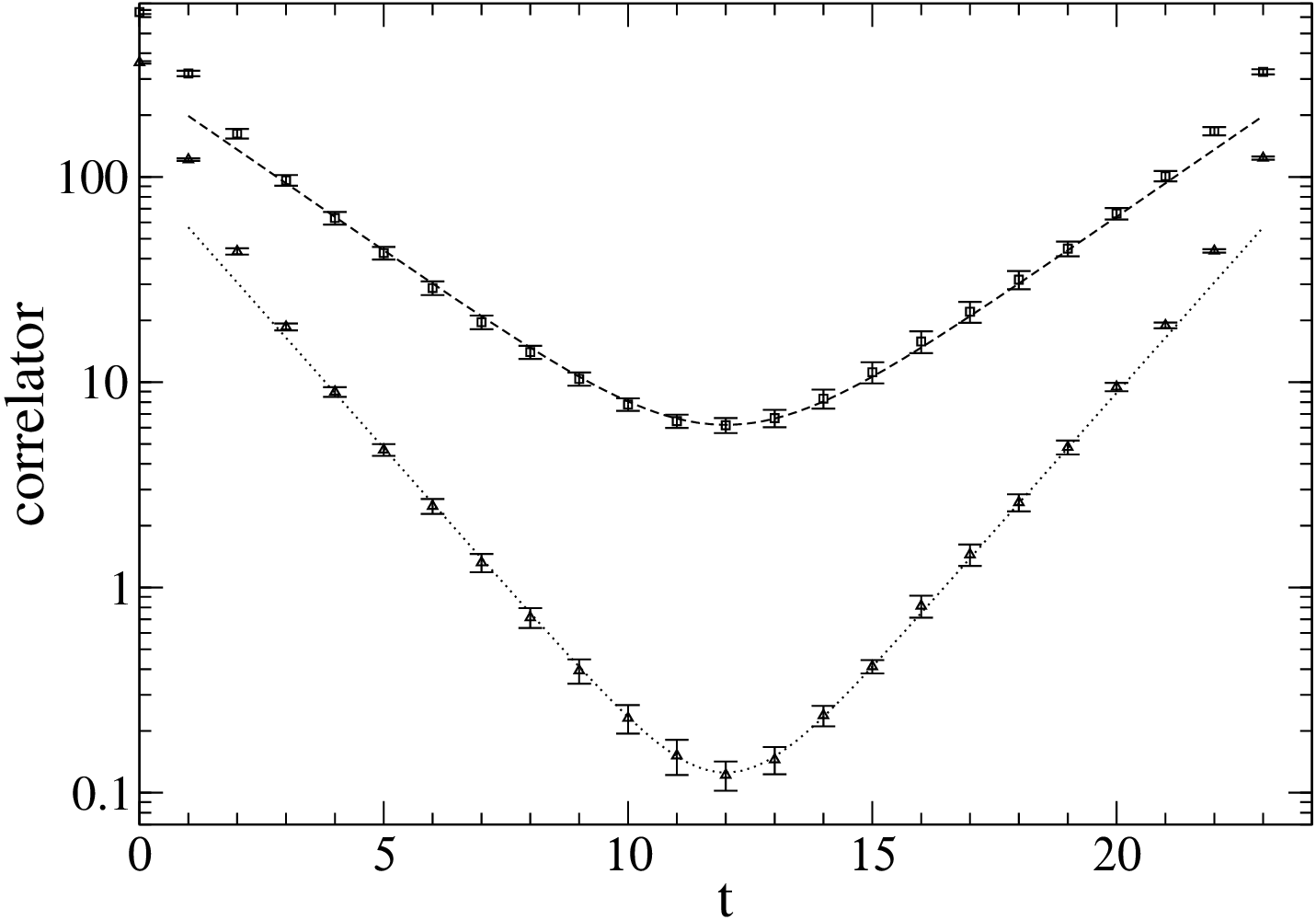}
\caption{
Correlation function for run (b), for the pseudoscalar (top) and the
vector (bottom) meson. The curves give the results of fits to the $\cosh$-behavior
as discussed in the text.}
\label{fig:pirho}
\end{figure}

\begin{table}[t]
\caption{Meson masses} 
\label{tab:masses}
\begin{ruledtabular}
\begin{tabular}{rllllll}
\#&   $aM_\pi$  & $aM_\rho$  &  $M_\pi/M_\rho$ &
   $M_\pi$[MeV]  & $M_\rho$[MeV]  \\
\hline
a&   0.292(10)&0.535(35)&0.55(5)  & 501(44)  &  918(109) \\
b&   0.378(8) &0.619(30)&0.61(4)  & 597(42)  &  977(96) \\
c&   0.326(18)&0.502(21)&0.65(6)  & 534(48)  &  823(62) \\
d&   0.431(8) &0.626(18)&0.69(3)  & 657(16)  &  954(33) \\
\end{tabular}
\end{ruledtabular}
\end{table}
\subsection{Meson masses}\label{sec:mesonmasses}

For the determination of the correlation function we used point quark
sources and Jacobi-smeared sinks. Our definition and notation for the
Jacobi smearing followed the quenched studies in \cite{BuGaGl05a,GaHuLa05a}.
We used the narrow smearing distribution. Smearing the sink crucially
improves the correlator signal quality and is almost as
efficient as smearing source and sink. 

The meson interpolating field were
the usual ones:
\begin{eqnarray}
\textrm{Pseudoscalar:}\quad P&=&\overline{d}\, \gamma_5 \,u\;,\\
                          A_4&=&\overline{d}\, \gamma_5 \gamma_4\, u\;,\\
\textrm{Vector:}    \quad V_k&=&\overline{d}\, \gamma_k \,u\;,
\end{eqnarray}
where $u$ and $d$ denote the up- and down quark fields; $A_4$ is the temporal component of the axial vector, 
which also couples to the pion.

We computed the correlation functions
\begin{eqnarray} 
\label{eq:corrPP}
&&\langle  P(\vec p =0,t) P(0) \rangle\;, \\
&&\langle  A_4(\vec p =0,t) A_4(0) \rangle\;,\\
&&\langle  V_i(\vec p =0,t) V_i(0) \rangle\;.
\end{eqnarray}

The result (see, e.g., Fig.\ \ref{fig:pirho}) was  then fitted to
\begin{equation}
C(t)= D(M)\,\left(\mathrm{e}^{-M \,t}\pm\mathrm{e}^{-M\,(T-t)}\right)\;.
\end{equation}
The masses thus may be derived from the exponential decay and other low energy
parameters from its coefficient.  

We discuss here only results for the large lattice size $12^3\times 24$. There
the fits were done in the range $(t_a,\,t_b)=(6,18)$. Table \ref{tab:masses}
summarizes our results for the masses. All error bars have been determined with
the jack-knife method. We also checked that $\chi^2/d.o.f.$ for  all the fits
were $<1$. The results for the $\langle  A_4 A_4\rangle$ correlator were
consistent with that for the $\langle P P\rangle$ correlation function, but
with slightly larger  statistical errors. In the table we therefore quote the
pion mass obtained from the $\langle P P\rangle$ correlator.

\subsection{AWI mass and pion decay constant}\label{sec:AWImass}

The axial Ward identity (AWI) allows one to define the renormalized quark 
mass through the asymptotic behavior of the ratio
\begin{equation}
\label{eq:ratDAXPX}
\frac{Z_A}{Z_P}\,
\frac{\langle \,\partial_t A_4(\vec p=\vec 0,t)\, X(0)\,\rangle}
{\langle\, P(\vec p=\vec 0,t)\,X(0) \,\rangle}
= Z_m\,2\,m=2\,m^{(r)}\;,
\end{equation}
where $X$ is any interpolator coupling to the pion and
$Z_A$, $Z_P$ and $Z_m$ denote the renormalization factors relating the 
$\overline{\mbox{MS}}$-scheme at a scale of 2 \textrm{GeV}. These have been
calculated for the quenched case at several values of the lattice spacing and 
came out close to 1 \cite{GaGoHu04}. We do not know the value for the dynamical case
but for the mass values presented we expect it to be close to the quenched one.
We therefore compute the ratio
\begin{equation}
\label{eq:ratDAPPP}
\frac{\langle \,\partial_t A_4(\vec p=\vec 0,t)\, P(0)\,\rangle}
{\langle\, P(\vec p=\vec 0,t)\,P(0) \,\rangle}
\equiv 2\,m_\srm{AWI}\;,
\end{equation}
defining the so-called AWI-mass.

We measure
\begin{equation}
\langle  A_4(\vec p =0,t) P(0) \rangle\;,
\end{equation}
in order to construct
\begin{equation}
\langle  \partial_t A_4(\vec p =0,t) P(0) \rangle \;.
\end{equation}
Ratios involving the lattice derivative $\partial_t A_4$ depend on the way the
derivative is taken. Numerical derivatives are always based on assumptions on
the interpolating function. Usual simple 2- or 3-point formulas assume
polynomials as interpolating functions. We can do better by utilizing the
information on the expected $\sinh$-dependence. In fact, we may use this
function for local 3-point $(t-1,\, t,\, t+1)$ interpolation and get the
derivative at $t$ therefrom.  We cannot use correlators like 
$\langle X(t)\,\partial_t A_4(0)\rangle$  since the source is fixed to the
time slice $t=0$ and thus we cannot construct the  lattice derivative there. 

Eq. (\ref{eq:ratDAPPP}) assumes interpolating fields with
point quark sources. 
The Jacobi-smearing of the quark sinks introduces a normalization factor
relative to point sinks. The factors can be obtained from the asymptotic (large $t$)
ratios of, e.g., 
\begin{equation}
c_P=\frac{\langle  P(t) P\rangle}{\langle P_{s}(t) P\rangle}\;,\quad
c_A=\frac{\langle  A_4(t) P\rangle}{\langle A_{4,s}(t) P\rangle}\;,
\end{equation}
where the index $s$ denotes the interpolator built from smeared sources.
We find values of $c_P/c_A$ around 0.6 indicating that the smearing of the quarks
affects the two operators differently, which may be understood from
their different Dirac content and ``wave function''. 
We did check that the results agree with that derived directly from correlators based
on point-like quark sinks, only the error bars are slightly larger in the latter case. 
All numbers given refer to the normalization for operators built from
point-like quark sources and sinks.

\begin{figure}[t]
\includegraphics*[width=8.3cm,clip]{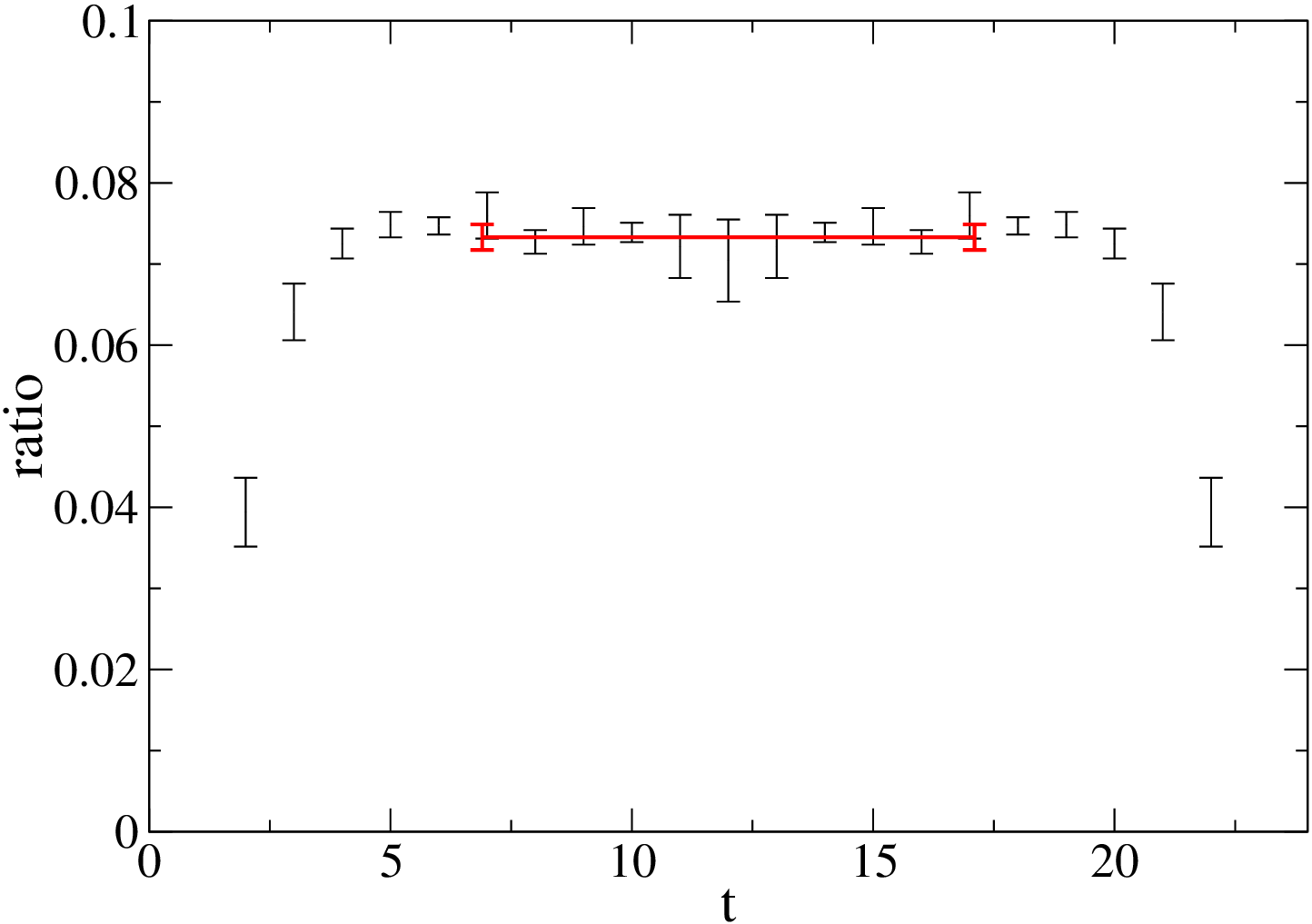}
\caption{The ratio (\ref{eq:ratDAPPP}) for $\beta=5.2$, $a\,m=.03$. 
The plateau fit range is indicated.}
\label{fig:mawi}
\end{figure}

Taking this into account we find the AWI-mass from plateau values like
that shown in Fig.\ \ref{fig:mawi}. The final average was taken in the same interval
as was used for the mass analysis and the error was again computed with the jack-knife method.
The results are given in Table \ref{tab:phys_data}.

In full, renormalized QCD the Gell-Mann--Oakes--Renner (GMOR) relation relates
the pion and quark masses:
\begin{equation}\label{eq:GMOR}
f_\pi^2\,M_\pi^2=- 2\, m\, \Sigma\;.
\end{equation}
Here two flavors of mass-degenerate quarks are assumed.
The quark mass and the condensate (contribution per flavor d.o.f.)
are renormalization scheme dependent and have to be given in, e.g.,
the $\overline{\mbox{MS}}$-scheme. Since the AWI-mass is proportional
to the renormalized quark mass, this linear relationship may hold.
Indeed, in lattice calculations surprisingly linear behavior has been found.

\begin{table}[bt]
\caption{Results for the AWI-mass in lattice units, and
the AWI-mass, $M_\pi$ and the pion decay constant in units of
the Sommer-parameter $r_0$ (which is usually assumed to be 0.5 fm).}
\label{tab:phys_data}
\begin{ruledtabular}
\begin{tabular}{cllll}
\#&  $am_\srm{AWI}$ & $r_0^2M_{\pi}^2$ & $r_0\,m_{\srm AWI}$ & $r_0\,f_{\pi}$  \\  
\hline
a & 0.025(1)& 1.62(28)         & 0.103(9) & 0.237(44)  \\
b & 0.037(1)& 2.29(33)         & 0.147(10)& 0.321(45)  \\
c & 0.037(2)& 1.84(33)         & 0.154(13)& 0.314(44)  \\
d & 0.050(1)& 2.78(14)         & 0.195(6) & 0.281(26)  
\end{tabular}
\end{ruledtabular}
\end{table}

\begin{figure}[t]
\includegraphics*[width=8.3cm,clip]{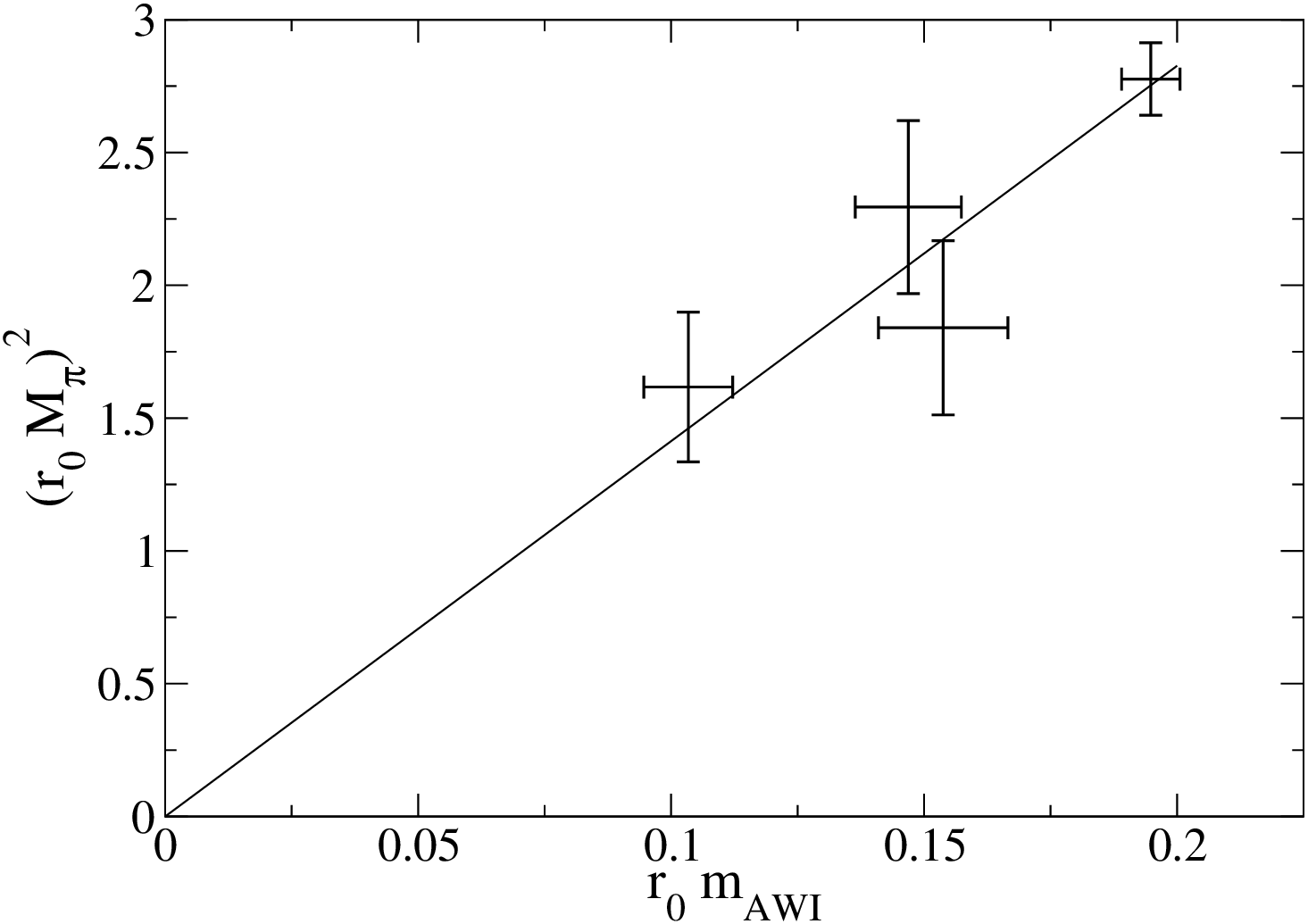}
\caption{
The pion mass squared vs the AWI-mass in units of the Sommer parameter $r_0$
for all 4 data sets (a--d). The fitted line corresponds to the lowest order
chiral perturbation theory behavior, i.e., to the GMOR relation (\ref{eq:GMOR}).}
\label{fig:GMOR}
\end{figure}

In Fig. \ref{fig:GMOR}  we plot our results for $M_\pi^2$ and $m_\srm{AWI}$
for all four runs. Within the errors the results are compatible with the expected
linear dependence. Neglecting the renormalization factors and taking the 
experimental value 92~MeV for the pion decay constant the slope, via 
(\ref{eq:GMOR}) corresponds to a value for the condensate of 
$\Sigma=(288(8)\;\textrm{MeV})^3$. The errors are purely statistical, from the fit
to the straight line, neglecting possible higher order chiral perturbation theory
contributions.

\begin{figure}[t]
\includegraphics*[width=8.3cm, clip]{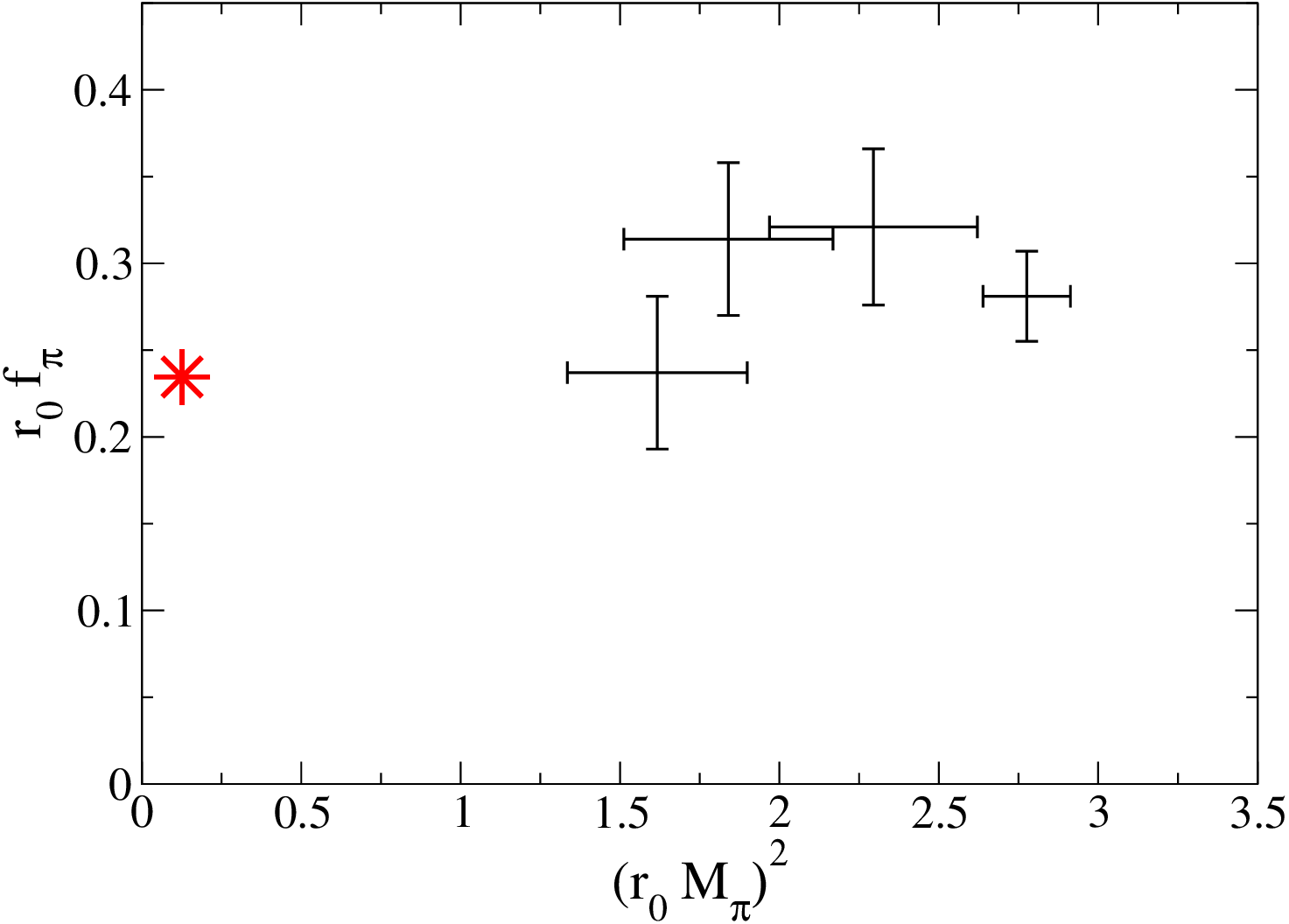}
\caption{$f_{\pi}$ vs pion mass squared in units of the Sommer parameter $r_0$
for all 4 data sets (a--d). The $\ast$ denotes the experimental value.}
\label{fig:fpi}
\end{figure}

We can now go ahead and obtain the pion decay constant $f_{\pi}$. It is related 
to the coefficient of the $ \langle A_4 A_4\rangle$ correlator with point source and sink 
and may be extracted through its asymptotic (large $t$) behavior:
\begin{equation}\label{fpi}
Z_A^2 \langle A_4(\vec p=\vec 0,t)A_4(0)\rangle
\stackrel{\textrm{\scriptsize large $t$}}{\longrightarrow} M_{\pi}f_{\pi}^2e^{-M_{\pi}t}\;.
\end{equation}

Again we assume $Z_A\approx 1$; in Fig.\ \ref{fig:fpi} we show the results, which are compatible with the
experimental values. For this measurement we actually used the point - point
correlation function. Table \ref{tab:phys_data} summarizes our results in
units of the Sommer parameter.

\section{Summary/Conclusion}

We have implemented the Chirally Improved Dirac operator, obeying the  GW 
condition to a good approximation, in a simulation with two species of
mass-degenerate light quarks. In our simulations on $12^3\times 24$-lattices the
lattice spacing was as small as 0.115 fm. The quark masses reached so far
correspond to a $M_\pi/M_\rho$ ratio of 0.55.

We consider this as a first step towards the final goal to approach realistic
quark masses. Nice features of the $\DCI$, which were observed in quenched 
simulations, like the spectrum following the GW circle, appear to hold in the
dynamical case as well.  The good consistency of the results with experimental
numbers seems to indicate, that the renormalization factors are close to 1 (as
they have been found in the quenched case). We also find no evidence for
spurious zero modes and we therefore do not expect serious
problems to go down further in  the quark mass. 

The results are encouraging and demonstrate that the CI operator is well suited for
dynamical fermion simulations.

\begin{acknowledgments}
We want to thank Christof Gattringer for helpful discussions. Support by Fonds
zur F\"orderung der  Wissenschaftlichen Forschung in \"Osterreich (FWF project
P16310-N08) is gratefully acknowledged. P.M. thanks the FWF for granting a
Lise-Meitner Fellowship (FWF project M870-N08). The calculation have been
computed on the Hitachi SR8000 at the Leibniz  Rechenzentrum in Munich and at
the Sun Fire V20z cluster of the computer center of
Karl-Franzens-Universit\"at, Graz, and we want to thank both institutions for
support.
\end{acknowledgments}


\end{document}